# Two-point functions of four-dimensional simplicial quantum gravity

W. Beirl, H. Markum and J. Riedler*

Institut für Kernphysik, Technische Universität Wien, A-1040 Vienna, Austria

We investigate the interaction mechanism of pure quantum gravity in Regge discretization. We compute volume-volume and link-link correlation functions. In a preliminary analysis the forces turn out to be of Yukawa type, at least on our finite lattice being away from the continuum limit.

## 1. INTRODUCTION

Regge calculus gives a description of General Relativity using simplicial approximations of space-time geometries [1]. A Regge skeleton consisting of a set of four-simplices is completely specified by its incidence matrix and by the link lengths. Here we use skeletons with a fixed incidence matrix restricting the topology to a four-torus and consider the link lengths as the dynamical variables of the theory. To define the Euclidean path integral one must choose a functional measure which ensures that the simplices satisfy the triangle inequalities and their higher dimensional analogs.

The functional integral with the Regge-Einstein action is

$$Z = \int \prod_l \frac{dq_l}{q_l^{1-\sigma}} \mathcal{F}[q_l] e^{\beta \sum_t A_t \delta_t - \lambda \sum_s V_s} , \qquad (1)$$

where triangle areas $A_t$, deficit angles $\delta_t$, and four-simplex volumes $V_s$ are calculated from the squared link lengths $q_l$. The function $\mathcal{F}$ is equal to one for Euclidean configurations and zero otherwise and the parameter $\sigma$ allows to consider different types of measure. Scaling arguments suggest to set the cosmological constant $\lambda = \sigma$ [2].

The phase structure of the theory (1) has already been studied extensively [2–4]. We mention only that in spite of an unbounded gravitational action the system develops a well-defined phase with small negative average curvature for couplings $\beta < \beta_c$. In order to facilitate numerical simulations a lower limit $f = 10^{-4} \leq \phi_s$ has been applied to the fatness of every four-simplex [4,5]

$$\phi_s \sim \frac{V_s^2}{\max_{l \in s}(q_l^4)} . \qquad (2)$$

## 2. CORRELATION FUNCTIONS

To understand the interaction mechanism of a theory one has to study correlation functions. In case of the Regge approach the observables $\mathcal{O}(x)$ are associated with a vertex $x$ and the corresponding expectation values

$$G(d) = \langle \mathcal{O}(x) \mathcal{O}(y) \rangle \qquad (3)$$

should be measured at two points separated by the geodesic distance $d = |x - y|$ [2]. As observables one can consider quantities like the local volume, the local curvature, the link lengths or the deficit angles. Although not all of these objects are diffeomorphism invariant in the continuum their computation yields insight into the behavior of the Regge skeleton.

For a first study we investigated connected volume-volume correlation functions

$$G_V(d) = \langle V(0) V(d) \rangle_c , \qquad (4)$$

where $V$ denotes the hypercubic volume element consisting of 24 four-simplices associated with each vertex. We performed MC simulations for several $\beta$ values on lattices with $3^3 \times 8$ and $4^3 \times 16$ vertices applying 100k - 300k and 30k - 70k sweeps, respectively. For asymmetric lattices it seems justified to use the index $n$ labeling the vertices in the long direction instead of the geodesic distance $d$. This avoids the time consuming determination of geodesic distances on the fluctuating simplicial lattice.

*Supported in part by "Fonds zur Förderung der wissenschaftlichen Forschung" under Contract P9522-PHY.

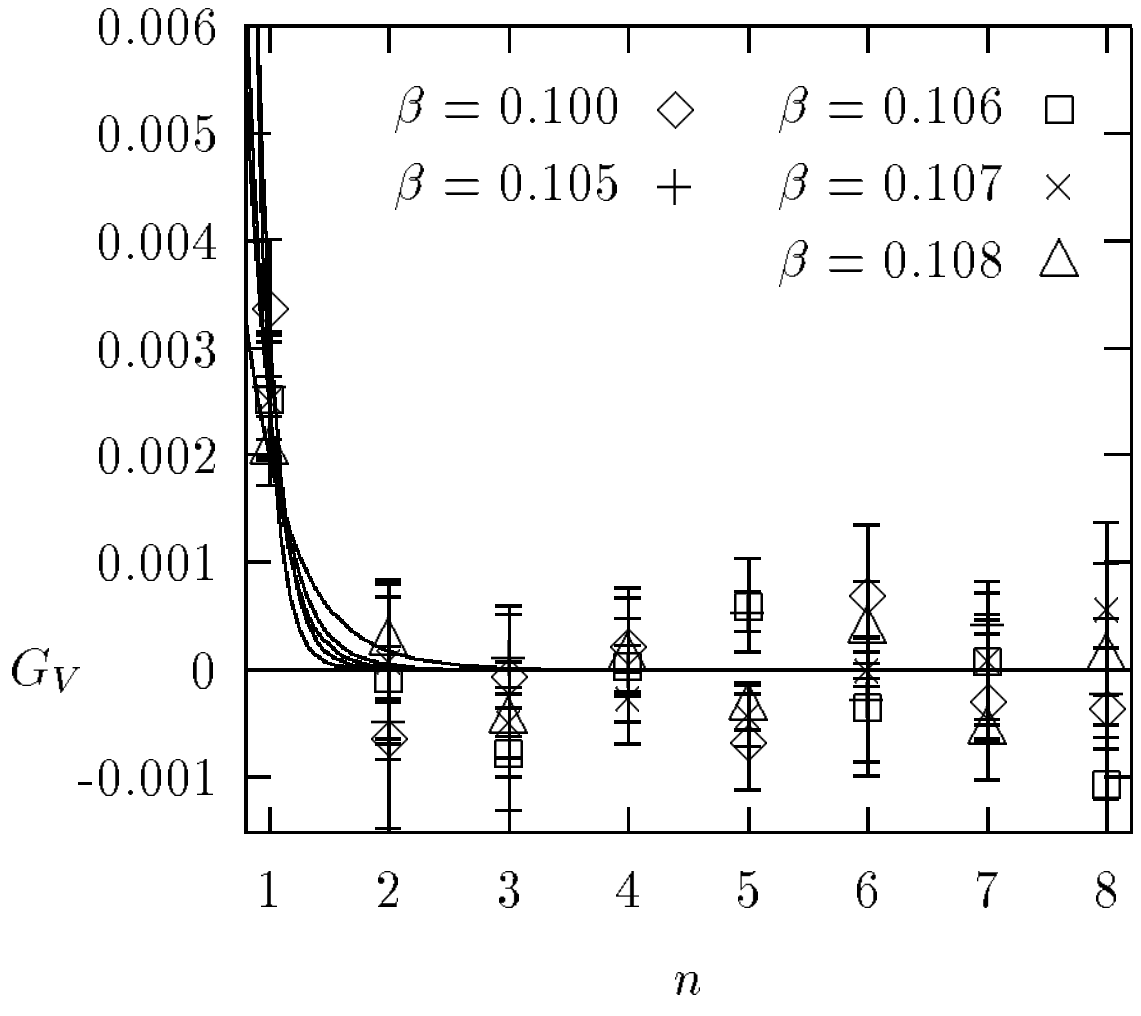

Figure 1. Volume correlations $G_V(n)$ for several couplings $\beta$ and $\lambda = \sigma = 1$ on a $4^3 \times 16$ lattice. The fits suggest a Yukawa like theory.

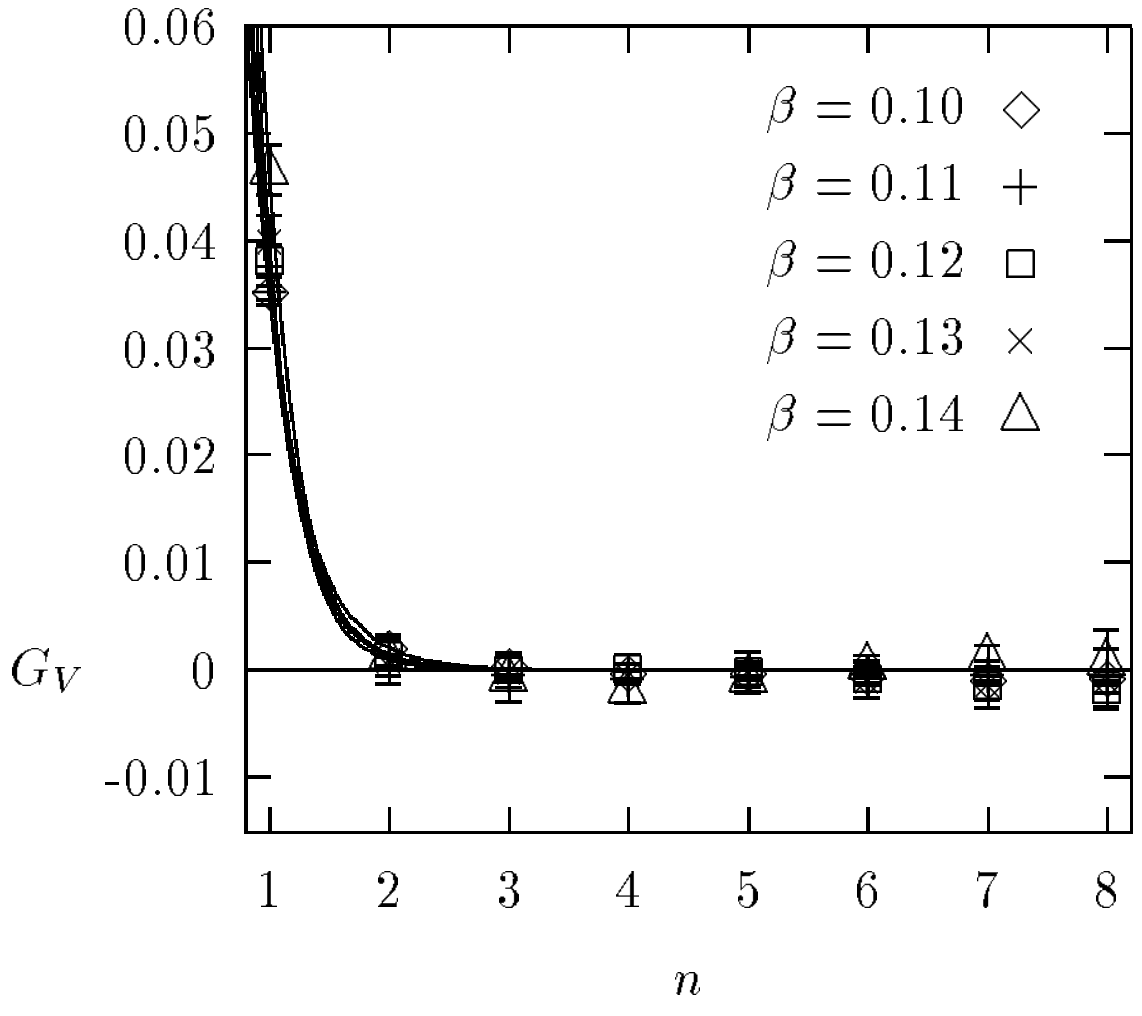

Figure 2. Volume correlations $G_V(n)$ for several couplings $\beta$ and $\lambda = \sigma = 0.1$ on a $4^3 \times 16$ lattice. Screening masses decrease with decreasing cosmological constant.

In Fig. 1 we present our results for the volume correlations in the case of a uniform measure and cosmological constant $\lambda = 1$. We find a fast decay of the correlators for all investigated $\beta$ values near the critical coupling $\beta_c \gtrsim 0.108$. In a preliminary analysis of the data we only tried an exponential decay without any power law term. We mention that the corresponding mass lies in the range $2 < m < 8$ in inverse lattice units. The cosmological term is expected to lead to a non-vanishing screening mass, $m \sim \lambda$ [6]. Therefore, we lowered $\lambda = \sigma$ to 0.1 approaching a scale invariant measure. In Fig. 2 we display the volume correlators for several $\beta$ values near $\beta_c \gtrsim 0.14$. Note the difference in the scale by a factor of 10. A fit to an exponential decay yields an effective mass $2 < m < 4$ being smaller than in the $\lambda = 1$ case as naively expected. Since it was demonstrated in earlier work that one-point expectation values are practically identical in the range $0 \leq \sigma \lesssim 1$ [4] it seems reasonable to compare directly the effect of the cosmological constant.

To gain further information about the geometry of the simplicial lattice we computed the $15 \times 15$ correlation matrix between the (squared) link lengths

$$Q_{ab}(d) = \langle q_a(0) q_b(d) \rangle_c \qquad (5)$$

and investigated its eigenvalue spectrum $\Lambda_a(d)$. The indices $a, b$ label the different types of edges within a hypercube along the indexing scheme of Ref. [7]. Some care must be taken to interpret $Q_{ab}(d)$ as a propagator. From perturbative calculations considering small fluctuations around flat space [7] it is known that the free inverse propagator on the lattice has zero eigenvalues (as well as in the continuum). One can argue that these zero modes have no extensions to a general curved lattice since diffeomorphism invariance is lost. The formulation of gauge transformations and gauge fixing are merely worked out for lattice gravity. Besides, the construction of physical two-point functions has been criticized even for continuum quantum gravity [8] but the fact that the lattice theory is formulated in a coordinate invariant manner at least addresses one of these objections [2].

In Fig. 3 we draw our results for the largest eigenvalue of the link correlations for uniform measure with $\lambda = 1$. Because our data shows large fluctuations we simply fitted an exponential without power law behavior and got for the mass $0.3 < m < 0.7$. Fig. 4 gives similar link correlations for an approximately scale invariant measure with small cosmological term $\lambda = \sigma = 0.1$. An exponential fit yields $1 < m < 3$.

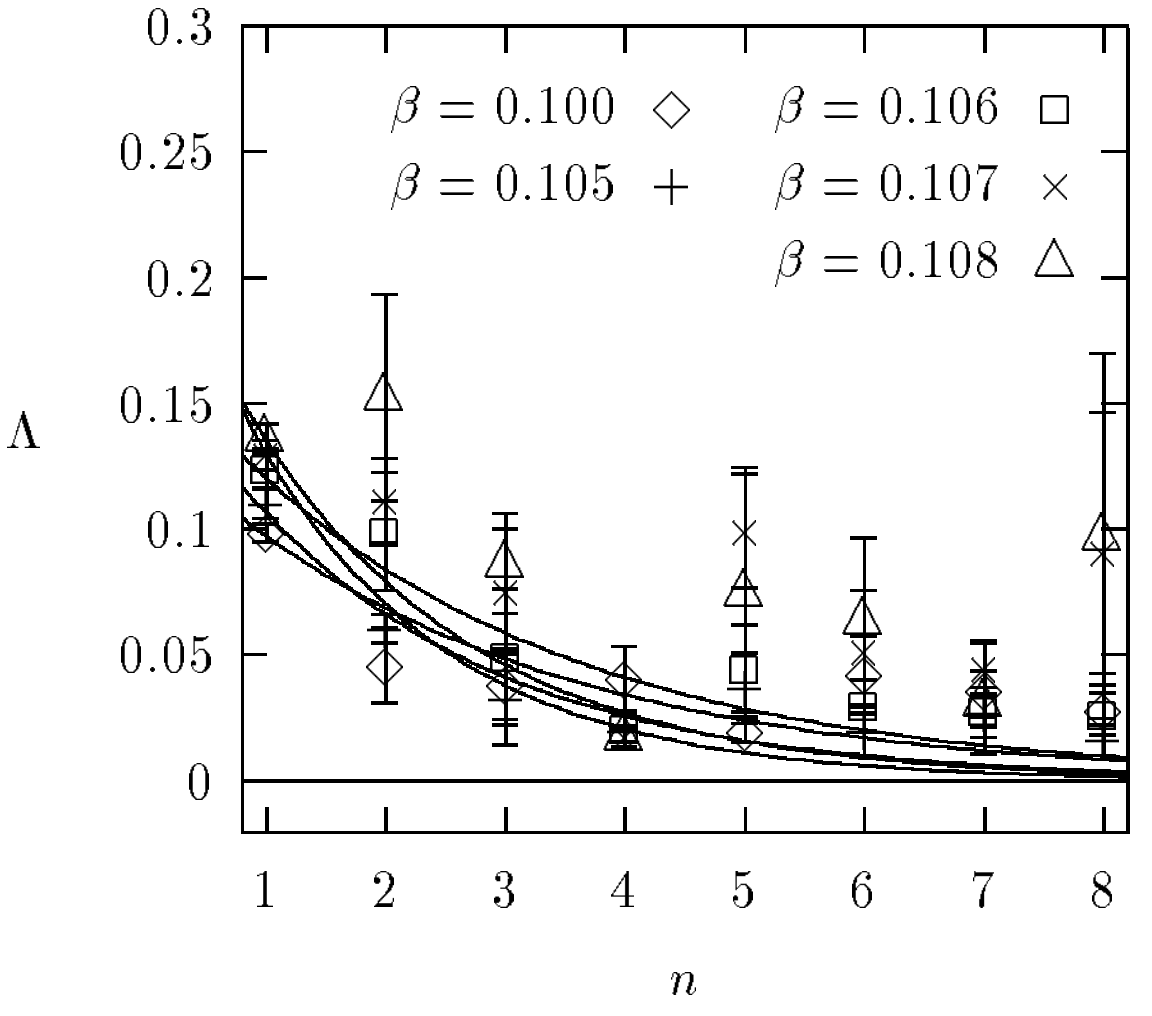

Figure 3. Largest eigenvalue $\Lambda(n)$ of the link-correlation matrix $Q_{ab}$ for $\lambda = \sigma = 1$ on a $4^3 \times 16$ lattice and the same couplings $\beta$ as in Fig. 1.

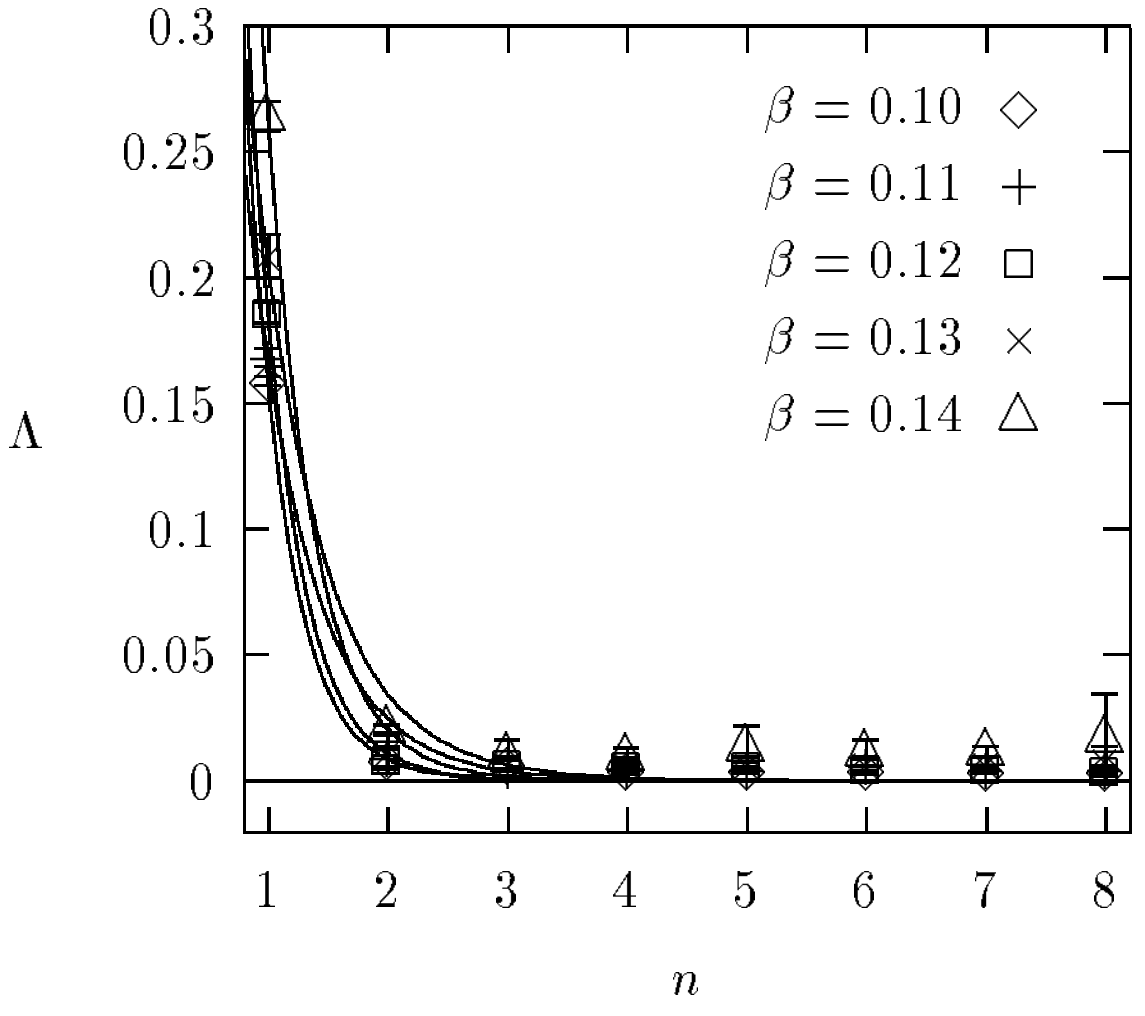

Figure 4. Largest eigenvalue $\Lambda(n)$ of the link-correlation matrix $Q_{ab}$ for $\lambda = \sigma = 0.1$ on a $4^3 \times 16$ lattice and the same couplings $\beta$ as in Fig. 2.

## 3. CONCLUSION

This work contains a first computation of two-point functions for four-dimensional simplicial quantum gravity. The preliminary results should be understood as a status report of a project considering different types of correlations. From the viewpoint of weak-field approximation with all the problems of a continuation to non-perturbative quantum gravity one might expect the curvature correlator as a very promising candidate to carry the quantum numbers of the graviton [2]. Our exploratory data analysis for the theory at finite lattice spacing indicates Yukawa forces with relatively large masses. The crucial question is the behavior of the mass spectrum in the continuum limit.